\documentstyle[12pt]{article}
\def\ZZZ{{\hbox{ Z\kern-1.6mm Z}}}
\def\zzz{{\hbox{ z\kern-1mm z}}}

\newcommand{\bz}{\bar z}
\newcommand{\bw}{\bar w}
\newcommand{\bJ}{\bar J}
\newcommand{\bA}{{\bf A}}
\newcommand{\bB}{{\bf B}}
\newcommand{\bF}{{\bf F}}

\newcommand{\bv}{\bar v}

\newcommand{\LL}{{\cal L}}

\newcommand{\wh}{\widehat}

\newcommand{\SSS}{{\cal S}}

\newcommand{\be}{\begin{equation}}
\newcommand{\ee}{\end{equation}}
\newcommand{\ben}{\begin{eqnarray}\displaystyle}
\newcommand{\een}{\end{eqnarray}}
\newcommand{\refb}[1]{(\ref{#1})}
\newcommand{\p}{\partial}

\def\one{{\hbox{ 1\kern-.8mm l}}}
\def\zero{{\hbox{ 0\kern-1.5mm 0}}}

\begin{document}

{}~
{}~

\hfill\vbox{\hbox{arXiv:0705.0735}}\break

\vskip .6cm

\medskip

\baselineskip 20pt 

\begin{center}

{\Large \bf
$AdS_3$, Black Holes and Higher Derivative Corrections}

\end{center}

\vskip .6cm

\vspace*{4.0ex}

\centerline{\large \rm Justin R. David,
Bindusar Sahoo and Ashoke Sen}

\vspace*{4.0ex}

\centerline{\large \it Harish-Chandra Research Institute}

\centerline{\large \it  Chhatnag Road, Jhusi,
Allahabad 211019, INDIA}

\vspace*{1.0ex}

\centerline{E-mail:  justin, bindusar, sen@mri.ernet.in}

\vspace*{5.0ex}

\centerline{\bf Abstract} \bigskip

Using AdS/CFT correspondence and the Euclidean action formalism
for black hole entropy Kraus and Larsen have argued that the
entropy of a BTZ black hole in three dimensional supergravity
with (0,4) supersymmetry does not receive any
correction from  higher derivative terms in the action. 
We argue that as a consequence of AdS/CFT
correspondence the action of a
three dimensional supergravity 
with (0,4) supersymmetry cannot receive any higher
derivative correction except for those which can be removed by
field redefinition. The non-renormalization of the entropy then
follows as a consequence of this and the invariance of Wald's
formula under a field redefinition.

\vfill \eject

\baselineskip 18pt



BTZ solution describes a black hole in three 
dimensional theory of gravity with negative cosmological 
constant\cite{9204099} and often
appears as a factor in the near horizon geometry of
higher dimensional black holes in string theory\cite{9712251}.
Furthermore
the entropy of a BTZ black hole has a remarkable similarity to the
Cardy formula
for the degeneracy of states 
in the two dimensional conformal field theory\cite{brown}.
For these
reasons computation of the entropy of BTZ black holes has been an
important problem,
both in three dimensional
theories of gravity and also in string theory. 
Initial studies involved computing Bekenstein-Hawking formula
for BTZ black hole entropy in two derivative theories of gravity.
Later this was 
generalized to higher derivative theories of 
gravity\cite{9909061,0506176,0508218,0509148,0601228,0609074,
0611141}, 
where the lagrangian 
density contains arbitrary powers of Riemann tensor and its covariant 
derivatives as well as gravitational Chern-Simons 
terms\cite{djw}, both
in the Euclidean action formalism\cite{9804085} 
and in Wald's 
formalism\cite{9307038,9312023,9403028,9502009}.

While the above mentioned formalism tells us how to calculate the
entropy of a BTZ black hole for a given action with arbitrary higher
derivative terms, it does not tell us what these higher derivative
terms are. It was however argued by Kraus and 
Larsen\cite{0506176,0508218} using AdS/CFT correspondence that if the
three dimensional 
theory under consideration has at least (0,4) supersymmetry then
the entropy of a BTZ black hole of given mass and angular
momentum is determined completely in terms
of the coefficients of the gravitational and gauge Chern-Simons
terms in the action and hence does not receive any higher derivative
corrections. This result is somewhat surprising from the point of view
of the bulk theory, since for a given three dimensional theory of gravity
the entropy does have non-trivial dependence on all the higher derivative
terms. Thus one could wonder how the dependence of the entropy on these
higher derivative terms disappears by imposing the requirement of (0,4)
supersymmetry.

In this note we shall propose a simple explanation for this fact: (0,4)
supersymmetry prevents the addition of any higher derivative terms in the
supergravity action (except those which can be removed by field
redefinition and hence give an equivalent theory). 
Our argument is based 
on the following observation.
In AdS/CFT correspondence the boundary operators dual to the fields
in the supergravity multiplet are just the superconformal currents
associated with the (0,4) supersymmetry algebra. The
correlation functions of these operators in the boundary theory are
determined completely in terms of the central charges $c_L$, $c_R$ of
the left-moving Virasoro algebra and the right-moving super-Virasoro
algebra. Of these $c_R$ is related to the central charge $k_R$ of the
right-moving SU(2) currents which form the R-symmetry currents of the
super-Virasoro algebra and hence to the coefficient of the
Chern-Simons term of the associated SU(2) gauge fields in the bulk theory.
On the other hand $c_L-c_R$ is determined in terms of the coefficient of
the gravitational Chern-Simons term in the bulk theory. 
Thus the  knowledge
of the gauge and gravitational 
Chern-Simons terms in the bulk theory determines
all the correlation functions of (0,4) superconformal currents in the
boundary theory.
Since by AdS/CFT correspondence\cite{9711200}
these correlation functions in the
boundary theory determine completely the boundary S-matrix of the
supergravity fields\cite{9802109,9802150}, we conclude that the
coefficients of the gauge and gravitational Chern-Simons terms in
the bulk theory determine completely the boundary S-matrix elements
in this theory.

Now the boundary S-matrix elements are the
only perturbative observables of the bulk theory. Thus we expect that 
two different theories with the same boundary S-matrix must be 
equivalent.
(We shall elaborate on this later.)
Combining this with the observation made in
the last paragraph we see that
two different gravity theories, both with (0,4) supersymmetry
and the same coefficients of the gauge and gravitational Chern-Simons
terms, 
must be equivalent.
Put another way, once we have constructed a classical supergravity
theory with (0,4) supersymmetry and given coefficients of the
Chern-Simons terms, there cannot be any higher derivative corrections
to the action involving fields in the gravity supermultiplet
except for those which can be removed by
field redefinition.\footnote{Incidentally, since the correlation functions
of the superconformal currents in the boundary theory, expressed in terms 
of $k_R$ and $c_L$, are also not
affected by inclusion of non-planar graphs, it follows that the
supergravity action also does not receive any higher derivative
quantum 
corrections.
This however does not mean that BTZ black hole entropy is
protected from quantum corrections; the global geometry of BTZ black hole
is different from $AdS_3$, and due to this BTZ black hole entropy can receive
correction from terms in the action which cannot be written as integrals
of local Lagrangian density in three dimensions\cite{0611143}.}
The non-renormalization of the entropy of the BTZ
black hole then follows trivially from this fact.
The complete theory in the bulk
of course can have other matter multiplets whose action will receive
higher derivative corrections. However
since restriction to the
fields in the gravity supermultiplet
provides a consistent truncation of the theory,\footnote{In practice
this means that there is no term in the action that contains
a single power of a matter field and one or more powers of the 
supergravity
fields. In the CFT living on the boundary this is reflected in the 
fact that a correlation function involving a single primary field
other than identity
and
arbitrary number of superconformal currents vanishes.}
and since the BTZ
black hole is embedded in this subsector, its entropy will not be affected
by these additional higher derivative terms.

Our arguments will imply in particular that the five dimensional
supergravity action constructed in \cite{0611329}, after dimensional
reduction on a sphere, must be equivalent to the three
dimensional supergravity action
given in eq.\refb{ebosonic} below
with the precise relationship between
the various coefficients as given in eq.\refb{edefkr}. 
This in turn
would
explain why the analysis of the black hole entropy given in
\cite{0702072,0703087} (see also \cite{0703099}) 
agrees with the expected result. We should caution the
reader however that the field redefinition needed to arrive at the
action given in \refb{ebosonic} may not be invertible on all field
configurations. For example if we take a Chern-Simons action and
add to it
the usual kinetic term for a gauge field then the kinetic term can be
removed formally by a field redefinition. However the theory with
the kinetic term has an extra pole in the gauge field propagator
corresponding to a massive photon which is absent in the pure
Chern-Simons theory. This happens because the field redefinition
that takes us from the theory with the gauge kinetic term to pure
Chern-Simons theory is not invertible on the plane wave solution
describing the propagating massive 
photon. This however does not affect
our argument as long as the field redefinition is invertible on
slowly varying field configuration around the $AdS_3$ 
background. In
this context we note that such field redefinitions are carried out
routinely in string theory, {\it e.g.} in converting
a term in the gravitational
action quadratic in the Riemann tensor to the Gauss-Bonnet
combination. The former theory typically has extra poles in the
graviton propagator which are absent in the latter theory.

For completeness we shall now describe this unique (0,4) supergravity
action and compute
the entropy of a BTZ black hole from this action.
The action was
constructed in \cite{9904010,9904068}
by regarding the supergravity
as a gauge theory based on $SU(1,1)\times SU(1,1|2)$ 
algebra.\footnote{A different class of supergravity theories were
constructed in \cite{ach1,ach2,0610077}
based on the supergroup $Osp(p|2;R)\times Osp(q|2;R)$,
with
the supercharges transforming in the vector representation
of the R-symmetry group $SO(p)_L\times SO(q)_R$. Thus the
corresponding boundary theories will have a different superalgebra
and a different set of correlation functions, and our arguments
cannot be used to relate the bulk action of these theories to the action
given in \refb{esupercs}.
}
If $\Gamma_L$ and $\Gamma_R$ denote the (super-)connections
in the $SU(1,1)$ and $SU(1,1|2)$ algebras respectively, then the
action is taken to be a Chern-Simons action\cite{wit1} of the 
form:\ben \label{esupercs}
\SSS &=& -a_L\, \int d^3 x \left[ Tr( \Gamma_L \wedge 
d\Gamma_L
+ {2\over 3} \Gamma_L\wedge \Gamma_L \wedge \Gamma_L\right]
\nonumber \\
&& + a_R \, 
\int d^3 x \left[ Str( \Gamma_R \wedge 
d\Gamma_R
+ {2\over 3} \Gamma_R\wedge \Gamma_R 
\wedge \Gamma_R\right]\, ,
\een
where $a_L$ and $a_R$ are constants. 
Note that the usual metric degrees of freedom are 
encoded in the connections
$\Gamma_L$ and $\Gamma_R$\cite{wit2}.
Thus there is no obvious  way to add $SU(1,1)\times SU(1,1|2)$
invariant higher derivative terms in the
action involving the field strengths associated
with the connections $\Gamma_L$ and $\Gamma_R$. 
{}From this viewpoint also
it is natural that the supergravity action
does not receive any higher derivative corrections.

The bosonic fields of this theory include the metric $G_{MN}$
and an
SU(2) gauge field ${\bf A}_M$ ($0\le M\le 2$), represented as 
a $2\times 2$ anti-hermitian matrix valued
vector field.
After expressing the action
in the component notation and eliminating auxiliary fields using
their equations of motion we arrive at the action 
\be \label{ebosonic}
\SSS= \int d^3 x\left[\sqrt{-\det G} \left[ R
+ 2 m^2 \right] +  K \, \Omega_3(\wh\Gamma) 
- {k_R\over 4\pi}\, 
\epsilon^{MNP} Tr\left({\bf A}_M
\p_N {\bf A}_P + {2\over 3} {\bf A}_M {\bf A}_N {\bf A}_P\right)\right]
\ee
where
\be \label{ecarel}
{1\over m} = {1\over 2} (a_R+a_L), 
\qquad K = {1\over 2} (a_L-a_R)\, ,
\ee
\be \label{edefkr}
k_R = 4\pi a_R = 4\pi \, \left({1\over m} - K\right) \, ,
\ee
$\wh\Gamma$ is the Christoffel connection constructed out of 
the metric $G_{MN}$ and
\be \label{e2}
\Omega_3(\wh\Gamma) = \epsilon^{MNP} \left[{1\over 2}
\wh\Gamma^R_{MS} \p_N \wh\Gamma^S_{PR} + {1\over 3}
\wh\Gamma^R_{MS} \wh\Gamma^S_{NT} 
\wh\Gamma^T_{PR}\right]\, .
\ee
$\epsilon$ is the
totally anti-symmetric symbol with $\epsilon^{012}=1$. Note that although 
the action contains gravitational Chern-Simons term, there are no terms 
involving square of the Riemann tensor.

We shall now compute the entropy of a BTZ black hole in this
theory. We begin by
reviewing the result for BTZ black hole entropy in a general
higher derivative theory of gravity. For this it will be enough to keep
only the gravitational fields in the action, setting all other fields
to zero.
Let us consider a general gravitational action in three dimensions
of the form:
\be \label{e1}
S = \int d^3 x \left[ \sqrt{-\det G} \LL^{(3)}_0
+ K \, \Omega_3(\wh\Gamma)\right]\, ,
\ee
where
$\LL^{(3)}_0$ denotes an arbitrary scalar constructed out of the metric, 
the 
Riemann tensor and covariant derivatives of the Riemann tensor.
A general BTZ black hole in the three dimensional theory
is described by the metric:
\be \label{e16}
G_{MN} dx^M dx^N = -{
(\rho^2 - \rho_+^2) (\rho^2 - \rho_-^2)\over l^2 \rho^2} d\tau^2
+ {l^2 \rho^2 \over (\rho^2 - \rho_+^2) (\rho^2 - \rho_-^2)} d\rho^2
+ \rho^2 \left(dy - {\rho_+ \rho_-\over l \rho^2} d\tau\right)^2\, ,
\ee
where $l$, $\rho_+$ and $\rho_-$ are parameters labelling the
solution. Of these the parameters $\rho_\pm$ can be removed locally
by a coordinate transformation, so that any scalar combination of the
Riemann tensor and metric computed for this metric is a function of
the parameter $l$ only. We define
\be \label{edefhl}
h(l) = \LL_0^{(3)}\, ,
\ee
evaluated in the background \refb{e16}, and 
\be \label{edefgl}
g(l) = {\pi l^3\over 4}\, h(l)\, .
\ee
Then the following
results hold (see {\it e.g.} \cite{0601228}):
\begin{enumerate}
\item Equations of motion of the metric determines the value $l_0$ of
$l$ to be a solution to the equation
\be \label{ehsol}
g'(l_0)=0\, .
\ee
\item The entropy of a BTZ black hole with mass $M$ and angular
momentum $J$ is given by\footnote{It is worth emphasizing that
since under a field redefinition of the metric $l\to f(l)$ for some
function $f(l)$, $l_0$ is not invariant under a field redefinition. However
$g(l_0)$, being the value of the function $g(l)$ at its extremum, is
invariant under such a field redefinition.}
\be \label{eenbtz}
S_{BH} = 2\pi \sqrt{ c_L q_L\over 6} +  2\pi \sqrt{ c_R q_R\over 6}\, ,
\ee
where
\be \label{e34}
q_L={1\over 2} (M-J), \qquad q_R = {1\over 2} (M+J)\, ,
\ee
\be \label{e26}
c_L = 24\, \pi\, (C+K)\, , \qquad c_R = 
24\, \pi\, (C-K) \, ,
\ee
\be \label{ecdef}
C = -{1\over \pi} g(l_0)\, .
\ee
\item The parameters $\rho_\pm$ are related to $M$ and $J$ via the
relations
\be \label{em7}
M\pm J  =  {2\pi (C\mp K)\over l_0^2}
(\rho_+\pm \rho_-)^2\, .
\ee
\end{enumerate} 

We shall now apply these results to the action given in 
\refb{ebosonic}. 
We get
\be\label{ehlspec}
h(l) = (- 6 l^{-2} +2 m^2)\, ,
\qquad
g(l) = {\pi\over 4} l^3 \, (- 6 l^{-2} + 2 m^2)\, ,
\qquad
l_0 = {1\over m}\, ,
\qquad 
C = {1\over m}
\ee
and 
\be \label{ellcr}
c_L = 24\, \pi \, \left({1\over m} + K\right) = 24\, \pi \,
a_L, \qquad
c_R = 24\, \pi \, \left({1\over m} - K\right) = 24\, \pi \, a_R\, ,
\ee
where in \refb{ellcr} we have used \refb{ecarel}.
Using \refb{edefkr} we get
\be \label{ealar}
c_R = {6\, k_R}, \qquad c_L =  48\, \pi\, K + 6 \, k_R\, .
\ee
Eqs.\refb{eenbtz}, \refb{e34}, \refb{ealar} give the
desired expression for the entropy of a BTZ black hole in terms of
the coefficients of the gauge and gravitational Chern-Simons
terms. By our previous
argument addition of higher derivative terms do not change this 
result as long as they respect (0,4) supersymmetry.

Since the crux of our argument has been the relationship between
non-renormalization of the boundary S-matrix and the non-renormalization
of the classical action, we shall now elaborate on this by examining
how this works for the gauge sector of the theory. In this case the
Chern-Simons theory has equation of motion 
${\bf F}_{MN}=0$ where
\be\label{edeff}
\bF_{MN} \equiv \p_{[M}\bA_{N]} + [ \bA_M, \bA_N]
\ee
is the gauge field strength. Any additional gauge invariant
term in the action will involve the gauge field strength and hence
will vanish when ${\bf F}_{MN}=0$. A standard argument then
shows that such terms can be removed from the action using a
field redefinition.

We shall now see how the vanishing of the 
additional terms in the action for $\bF=0$
is related to the non-renormalization of the boundary S-matrix. 
For this we first review the computation of the boundary S-matrix
from pure Chern-Simons theory.
We begin by writing the Euclidean
$AdS_3$ metric in the Poincare patch
\be \label{epo1}
ds^2 = {l^2\over (x^0)^2} \left( (dx^0)^2 + (dx^1)^2 + (dx^2)^2\right)\, ,
\ee
and introduce complex coordinate  $z$, integration measure $d^2 z$ and 
the $\delta$-function $\delta^{(2)}(\vec z)$ as follows:
\be \label{ep01.5}
z = x^1 + i x^2\, , \qquad
d^2 z \equiv dx^1 dx^2 \, , \qquad \delta^{(2)}(\vec z) 
\equiv\delta(x^1)\delta(x^2)\, .
\ee
The gauge field action in the Euclidean space takes the 
form\footnote{We shall follow the same sign and normalization
convention as ref.\cite{0607138}. There is an apparent difference 
in the overall
sign of the Chern-Simons term, but this is related to the fact that
in the $x^0,x^1,x^2$ coordinate system the boundary of $AdS_3$
is at the lower limit of $x^0$ ($x^0=0$) and we have chosen 
$\epsilon^{012}>0$. As a result we need the $-$ sign in front of the
Chern-Simons term to ensure that the variation of the 
on-shell action depends
only on $\delta \bA_z$ and not on $\delta \bA_{\bz}$ at the
boundary.}
\be \label{epo2}
S_{gauge} = -i {k_R\over 4\pi} \, \int d^3 x \epsilon^{MNP}
Tr\left(\bA_M \p_N \bA_P +{2\over 3} \bA_M\bA_N\bA_P\right)
+{k_R\over 2\pi} \int d^2 z Tr(\bA_z \bA_{\bz})|_{x^0 =0}\, .
\ee
The last term in \refb{epo2}
is a boundary term needed to ensure the consistency
of the theory\cite{Elitzur,9904010,0607138}.
The effect of this term is that while computing the variation of the
on-shell action under a variation of the gauge fields, the result
depends only on $\delta \bA_z$ at the boundary $x^0=0$ and not on
$\delta \bA_{\bz}$. Thus while deriving the equations of motion  from
the action using a variational principle, we fix the boundary condition
only on $\bA_z$\cite{0607138}. 
Let us denote by $\vec z$ the pair $(z,\bz)$ and let
$I[\bA^{(0)}_z]$ denote the value of the Euclidean
action evaluated
for an on-shell field configuration subject to the boundary condition
\be \label{epo5.5}
\bA_z(x^0=0, \vec z) = \bA^{(0)}_z(\vec z)\, .
\ee 
Then according to
$AdS/CFT$ conjecture we have\cite{9802109,9802150}
\be \label{epo5}
\left\langle \bJ^{a_1}(\vec z _1) \cdots \bJ^{a_n}(\vec z_n) 
\right\rangle
= (i\pi)^n\, \left.
{\delta \over \delta A^{(0)a_1}_z(\vec z_1) \cdots 
\delta A^{(0)a_n}_z(\vec z_n)}
e^{-I[\bA^{(0)}_z]}\right|_{\bA^{(0)}_z(\vec z)=0}\, ,
\ee
where $\bJ^a(\vec z)$ are the $SU(2)$ currents of the CFT at the
boundary and the $A_M^a$ are defined through
\be \label{epauli}
\bA_M = {1\over 2} \, i\, \sigma^a \, A_M^a\, , 
\ee
$\sigma_a$ being the Pauli matrices.
Thus our task is to compute $I[\bA^{(0)}_z]$. For this we
first need to evaluate the gauge field configuration that satisfies
the equation of motion $\bF_{MN}=0$ and
the boundary condition 
\refb{epo5.5}.
This is given by
\be \label{eks2}
\bA_M(\vec z, x^0)\, dx^M = e^{-\Phi}\, d\, e^\Phi
= d\Phi + {1\over 2} (d\Phi \Phi - \Phi d\Phi) 
+\cdots\, 
\ee
where
\be \label{eks1}
\Phi(\vec z, x^0) =  
\int d^2 w K(\vec z, x^0; \vec w)
\, \bB^{(0)}_z(\vec w) \, ,
\ee
\be \label{epo7}
 K(\vec z, x^0; \vec w) = {1\over \pi}\,  
\left[ \frac {z -w}{(x^0)^2 + |z-w|^2}\right]\, ,
\ee
and $\bB^{(0)}_z$ 
is
chosen such that \refb{eks2} 
satisfies the boundary condition \refb{epo5.5}. 
Eq.\refb{epo7} gives
\be \label{epo8}
\lim_{x^0\to 0} \p_z 
K(\vec z, x^0; \vec w) = \delta^{(2)}(\vec z - \vec w)
\, ,
\ee
\be \label{epo8.1}
\lim_{x^0\to 0} \p_{\bz}K(\vec z, x^0; \vec w) = 
-{1\over \pi} \, {1\over (\bz
- \bw)^2}\, .
\ee
Using eqs.\refb{eks2}-\refb{epo8} we find that $\bA_z(x^0=0,\vec z)$
is equal to $\bB^{(0)}_z(\vec z)$ to first
order in an expansion in a power series in $\bB^{(0)}_z$. Thus to this
order \refb{epo5.5} is satisfied for
$\bB^{(0)}_z=\bA^{(0)}_z$. 
The higher order contributions
to $\bB^{(0)}_z$ can be obtained by 
iteratively solving eq.\refb{epo5.5} with the ansatz for $\bA_M$
given in \refb{eks2}-\refb{epo7}.
The result is
\ben \label{eks3}
\bB^{(0)}_z(\vec z) &=& \bA^{(0)}_z(\vec z)
+{1\over 2\pi} \int {d^2 w\over (\bz - \bw)} \, 
\left( \bA^{(0)}_z(\vec w)\bA^{(0)}_z(\vec z)-
\bA^{(0)}_z(\vec z)\bA^{(0)}_z(\vec w)\right)  +\cdots
\een
where $\cdots$ denote higher order terms. We can now substitute the
solution given in \refb{eks2}-\refb{eks3} 
into \refb{epo2} to evaluate the on-shell
action $I[\bA^{(0)}_z]$. 
Evaluation of the boundary contribution is straightforward. In
evaluating the contribution from the Chern-Simons term we first
use the equation of motion to express it as
\be \label{ecs1}
i {k_R\over 12\pi} \, \int d^3 x \epsilon^{MNP}
Tr\left(U^{-1}\p_M U U^{-1} \p_N U U^{-1}\p_P U\right)
\, ,
\ee
where $U=e^{\Phi}$. Defining $U_t=e^{t\Phi}$  and noting that
\be \label{ecs2}
{1\over 3}\epsilon^{MNP} \p_t Tr
\left(U_t^{-1}\p_M U_t U_t^{-1} \p_N U_t U_t^{-1}\p_P U_t\right)
= \epsilon^{MNP}\p_M 
\left(U_t^{-1}\p_t U_t U_t^{-1} \p_N U_t U_t^{-1}\p_P U_t\right)
\ee
and that $U_t^{-1}\p_t U_t=\Phi$,
we can express \refb{ecs1} as a pure boundary term
\be \label{ecs3}
-{k_R\over 2\pi}\, \int d^2 z\, \int_0^1 \, dt\, 
\left. Tr\left( \Phi \left[ U_t^{-1}\p_z U_t, U_t^{-1}
\p_{\bz} U_t\right]\right)\right|_{x^0=0}\, .
\ee
\refb{ecs3} can be evaluated by expanding the
integrand in a power series in $t$ and carrying out the $t$ integral
explicitly at every order.
The final 
result for the full action is:
\ben \label{eac1}
I[\bA^{(0)}_z] &=& 
{k_R\over 4\pi^2} \int d^2 z d^2 w \,  (\bz - \bw)^{-2} \, 
A^{(0)a}_z(\vec z)
\, A^{(0)a}_z(\vec w) \nonumber \\ &&
-{k_R\over 12\pi^3} \epsilon^{abc} \int d^2 z d^2 w d^2 v\,
(\bz - \bw)^{-1}(\bw-\bv)^{-1}(\bv-\bz)^{-1} \, A_z^{(0)a}(\vec z)
A_z^{(0)b}(\vec w) A_z^{(0)c}(\vec v)
\nonumber \\ && +\cdots \, .
\een
Eqs.\refb{epo5} and \refb{eac1} now give:
\be \label{eop3}
\left\langle \bJ^{a_1}(\bz _1) \bJ^{a_2}(\bz _2) \right\rangle
= {k_R\over 2} \, \delta_{a_1 a_2}\, (\bz _1-\bz _2)^{-2} \,  ,
\ee
\be \label{eop4}
\left\langle \bJ^{a_1}(\bz _1) \bJ^{a_2}(\bz _2) 
\bJ^{a_3}(\bz _3) \right\rangle
= - {i k_R\over 2} \, 
\epsilon^{a_1a_2 a_3} (\bz _1-\bz _2)^{-1} (\bz _2 - \bz _3)^{-1}
(\bz _3 - \bz _1)^{-1}\, ,
\ee
which are the expected conformal field theory correlation functions.
Following this procedure we can in principle calculate 
arbitrary correlation functions of the SU(2) currents.

Let us now consider the effect of including additional gauge invariant
terms in the action. Since such terms are functions of 
gauge field strength, the  solution \refb{eks2} satisfying 
$\bF_{MN}=0$ continues to be solution of the equations of motion
of the new theory. Furthermore the on-shell action is not modified
since all the additional terms  vanish when gauge field strength
vanishes. As a result the correlation functions of the currents
computed via \refb{epo5} also remains unchanged. Thus we see
that the vanishing of possible corrections to the correlators of
SU(2) currents is intimately related to the vanishing of the additional
terms in the action for an on-shell field configuration of the original
theory. The latter in turn implies that the additional terms in the action
can be removed by field redefinition.

We can now turn this argument around to 
see why the action
in the gravity sector is also
not renormalized. The non-renormalization
of the boundary S-matrix (which in turn follows from supersymmetry
relating the correlators of the currents and the stress tensor in the
CFT) implies
that any additional contribution to the action must vanish when original
equations of motion are satisfied. This in turn implies that such
additional terms can be removed by field redefinition. We emphasize
that supersymmetry is crucial for this argument. In absence
of supersymmetry relating the current correlators to the stress tensor
correlators there is no reason for the latter to be not renormalized. This
in turn would then imply that the effective action can receive corrections
which do not vanish when the original equations of motion are
satisfied. Hence such corrections
cannot be removed by field redefinition.

\end{document}